\documentclass[aip,reprint]{revtex4-1} 
 
\usepackage{graphicx} 
\usepackage{float} 
\usepackage{subfigure} 
\usepackage{epsfig} 
\usepackage{color}

\draft 
 
\begin{document} 
 
 
\newcommand\Alfven{Alfv\'en } 
\newcommand\Elsasser{Els$\ddot{\mbox{a}}$sser } 
\newcommand{\V}[1]{\mathbf{#1}}  
\newcommand{\xhat}{\mbox{$\hat{\mathbf{x}}$}}  
\newcommand{\yhat}{\mbox{$\hat{\mathbf{y}}$}}  
\newcommand{\zhat}{\mbox{$\hat{\mathbf{z}}$}}  
 
\title{\Alfven Wave Collisions, The Fundamental Building Block of Plasma Turbulence IV: Laboratory Experiment} 
 
 
 
\author{D. J. Drake} 
\email[]{djdrake@valdosta.edu} 
\affiliation{Department of Physics, Astronomy, and Geosciences, Valdosta State University, Valdosta, Georgia 31698} 
\author{J. W. R. Schroeder} 
\affiliation{Department of Physics and Astronomy, University of Iowa, Iowa City, Iowa 52242} 
\author{G. G. Howes} 
\affiliation{Department of Physics and Astronomy, University of Iowa, Iowa City, Iowa 52242} 
\author{C. A. Kletzing} 
\affiliation{Department of Physics and Astronomy, University of Iowa, Iowa City, Iowa 52242} 
\author{F. Skiff} 
\affiliation{Department of Physics and Astronomy, University of Iowa, Iowa City, Iowa 52242} 
\author{T. A. Carter} 
\affiliation{Department of Physics and Astronomy, University of California, Los Angeles, California 90095} 
\author{D. W. Auerbach} 
\affiliation{Department of Physics and Astronomy, University of California, Los Angeles, California 90095} 
 
 
\date{\today} 
 
\begin{abstract} 
Turbulence is a phenomenon found throughout space and astrophysical plasmas.  It plays an important role in solar coronal heating, acceleration of the solar wind, and heating of the interstellar medium. Turbulence in these regimes is dominated by \Alfven waves.  Most turbulence theories have been established using ideal plasma models, such as incompressible MHD.    However, there has been no experimental evidence to support the use of such models for weakly to moderately collisional plasmas which are relevant to various space and astrophysical plasma environments.  We present the first experiment to measure the nonlinear interaction between two counterpropagating \Alfven waves, which is the building block for astrophysical turbulence theories.  We present here four distinct tests that demonstrate conclusively that we have indeed measured the daughter \Alfven wave generated nonlinearly by a collision between counterpropagating \Alfven waves. 
\end{abstract} 
 
 
\maketitle 
 
\section{Introduction} 
 
Plasma turbulence is important for our understanding of the dynamics of various space and astrophysical plasma environments, including the heating of the interstellar medium, \cite{Spangler1998} acceleration of the solar wind,\cite{Verdini2007, Bingham2010} solar coronal heating,\cite{Cranmar2003} transport of energy and mass into Earth's magnetosphere,\cite{Lee1994, Sundkvist2005} and heat transport in galaxy clusters.\cite{Peterson2006, Lazarian2006}   Although these seem to be strikingly different environments, the turbulence in these plasmas is dominated by \Alfven waves, which are low frequency, large length scale waves.  This turbulent interaction arises when two counterpropagating \Alfven waves interact nonlinearly in the plasma medium.  This nonlinear interaction, often referred to as a wave-wave collision, is the central component of astrophysical turbulence and is responsible for the turbulent cascade of energy from large to small scales.\cite{Schekochihin2009}  In order to gain insight into this fundamental building block of astrophysical turbulence, experimental or observational measurements of the nonlinear interaction between two colliding \Alfven waves are essential.   
 
Turbulent fluctuations in the solar wind \cite{Mattaeus1982, Goldstein1995, Tu1995, Alexandrova2008,  Sahraoui2009, Spangler2011}  and the interstellar medium \cite{Duffet1975, Rickett1990, Gwinn1993, Scalo2004, Spangler2010} have been measured for several decades. However, these measurements are mostly used to study the effect of turbulence on the plasma environment. Although characterization of the effects of turbulence on these environments is important, they do little to explain the physical mechanisms comprising the turbulence.  More importantly, most of this data is limited by the fact that these are single-point measurements and, therefore, do not provide enough information about the 3-D structure of the turbulent fluctuations that leads to energy cascades from large to small scales. 
 
In contrast to observations by spacecraft missions and telescopes, the controlled nature of laboratory experiments allows for more detailed measurements of the small-scale nonlinear interactions between \Alfven waves.  In this paper, we provide detailed information about the experimental setup, experimental procedure, and data analysis of the first successful effort to confirm this nonlinear interaction, as outlined in \citet{Howes2012b}  In Sec. II, we briefly discuss the underlying theory.  A more detailed theory can be found in the three companion papers by \citet{Howes2013a}, hereafter Paper I, \citet{Howes2013b}, hereafter Paper II, and \citet{Howes2013c}, hereafter Paper III.  In Sec. III, we describe the experimental approach, with an emphasis on the two antennas used to produce the two counterpropagating \Alfven waves.  In Sec. IV, we present an analysis of the experimental results, directly comparing the measured nonlinear signal to the predicted results from the theory.    
 
\section{\Alfven wave turbulence theory} 
As shown in Paper I, the equations of incompressible MHD can be written in a symmetrized Els\"asser form,\citep{Elsasser1950} 
\begin{equation} 
\frac{\partial \V{z}^{\pm}}{\partial t}  
\mp \V{v}_A \cdot \nabla \V{z}^{\pm}  
=-  \V{z}^{\mp}\cdot \nabla \V{z}^{\pm} -\nabla P/\rho_0, 
\label{eq:elsasserpm} 
\end{equation} 
 
\begin{equation} 
\nabla \cdot  \V{z}^{\pm}= 0 
\label{eq:div0} 
\end{equation} 
where the magnetic field is decomposed into equilibrium and fluctuating parts $\V{B}=\V{B}_0+ \delta \V{B} $, $\V{v}_A=\V{B}_0/\sqrt{4 \pi\rho_0}$ is the \Alfven velocity due to the 
equilibrium field $\V{B}_0=B_0 \zhat$, $P$ is total pressure (thermal plus magnetic), $\rho_0$ is mass density, and $\V{z}^{\pm}(x,y,z,t) =\V{u} \pm \delta \V{B}/\sqrt{4 \pi \rho_0}$ are the Els\"asser fields given by the sum and difference of the velocity fluctuation $\V{u}$, and the magnetic field fluctuation $\delta \V{B}$ expressed in velocity units. This symmetrized Els\"asser form of the incompressible MHD equations lends itself to a particularly simple physical interpretation.  An \Alfven wave traveling down (up) the equilibrium magnetic field is represented by the Els\"asser field $\V{z}^{+}$ ($\V{z}^{-}$).  The second term on the left-hand side of Eq. (\ref{eq:elsasserpm}) is the \emph{linear term} representing the propagation of the Els\"asser fields along the mean magnetic field at the \Alfven speed, the first term on the right-hand side is the \emph{nonlinear term} representing the interaction between counterpropagating waves, and the second term on the right-hand side is a nonlinear term that ensures incompressibility.\cite{Montgomery1982, Goldreich1995, Howes2013a} 
 
As emphasized in Paper III, the mathematical properties of Eqs. (\ref{eq:elsasserpm}) and (\ref{eq:div0}) dictate that the fundamental building block of turbulence in an incompressible MHD 
plasma is the nonlinear interaction between perpendicularly polarized, counterpropagating \Alfven waves. Additionally, the strength of the nonlinear distortion of an \Alfven wave $\V{z}^{+}$ traveling down the equilibrium magnetic field is controlled by the amplitude of the counterpropagating \Alfven wave $\V{z}^{-}$ traveling up the magnetic 
field.  Therefore, to measure the nonlinear energy transfer in an \Alfven wave collision, one need only launch  a single \Alfven wave of large amplitude and then observe its effect on a 
counterpropagating \Alfven wave of smaller amplitude. 
 
Instrumental limitations on the amplitude of \Alfven waves launched in the experiment lead to a situation in which the nonlinear terms on the right-hand side of Eq. (\ref{eq:elsasserpm}) are small compared to the linear term on the left-hand side.  Therefore, the experimental dynamics falls into the weakly nonlinear limit, and we can exploit the developments in the theory of \emph{weak MHD turbulence}\cite{Iroshnikov1963, Kraichnan1965, Shebalin1983, Sridhar1994, Montgomery1995, Ng1996, Goldreich1997, Galtier2000, Lithwick2003} to optimize the experimental design, as discussed in detail in Paper III and briefly outlined in the remainder of this section.   
 
Consider the case of the nonlinear interaction between two counterpropagating plane \Alfven waves with wavevectors $\V{k}_1$ and $\V{k}_2$ and amplitudes $\delta B_1$ and $\delta B_2$. We want to design an experiment that will lead to a measurable nonlinear energy transfer to a third \emph{daughter} \Alfven wave with wavevector $\V{k}_3$. As shown in Paper I, the application of perturbation theory to obtain an asymptotic solution for the nonlinearly generated daughter \Alfven wave demonstrates that resonant three-wave interactions generate a daughter mode with an amplitude proportional to $(\delta B_1 \delta B_2/B_0^2)$, whereas resonant four-wave interactions nonlinearly generate modes with amplitudes proportional to $(\delta B_1^2 \delta B_2/B_0^3)$ or $(\delta B_1 \delta B_2^2/B_0^3)$.  Since instrumental limitations lead to \Alfven wave amplitudes that are always small compared to the equilibrium magnetic field, $\delta B_{1,2} \ll B_0$, it is desirable to design an experiment that will create a resonant three-wave interaction between the primary \Alfven waves. 
 
The theory of weak MHD turbulence demonstrates that, when averaged over an integral number of wave periods, resonant three-wave interactions must satisfy the resonance 
conditions\cite{Shebalin1983,Sridhar1994,Galtier2000}  
\begin{equation} 
\V{k}_1+ \V{k}_2 = \V{k}_3 \quad \mbox{ and } \quad \omega_1 + \omega_2 = \omega_3. 
\label{eq:constraints} 
\end{equation} 
Given the linear dispersion relation for \Alfven waves,
$\omega=|k_\parallel| v_A$, the only nontrivial solution to both
constraints in Eq. (\ref{eq:constraints}) therefore has either
$k_{\parallel 1}=0$ or $k_{\parallel 2}=0$.\cite{Shebalin1983} Thus,
as highlighted in Paper III, to obtain a nonzero, resonant three-wave
interaction between two counterpropagating \Alfven waves, it is
necessary to design an experiment such that the interacting waveform
of one of the \Alfven waves has a significant $k_{\parallel }=0$
component.\cite{Ng1996} 
This can be achieved if, over the length of the experiment in which
the two counterpropagating \Alfven waves interact, the wavepacket of
one of the \Alfven waves has a magnetic field perturbation is not
symmetric about $\delta B_\perp=0$.  In this case, the propagating
\Alfven wavepacket contains a nonzero $k_{\parallel }=0$
component that leads to a resonant three-wave interaction.

In the experiment described here, a Loop antenna generates a
low-frequency, large-amplitude \Alfven wave $\V{z}^{-}$ traveling in
the direction of the equilibrium magnetic field. This \Alfven wave
nonlinearly distorts a higher frequency, smaller amplitude \Alfven
wave $\V{z}^{+}$ that is launched by an Arbitrary Spatial Waveform
(ASW) antenna and travels opposite the direction of the equilibrium
magnetic field.  The design of the experiment achieves a significant
$k_{\parallel }=0$ component to the large-amplitude \Alfven wave
$\V{z}^{-}$ by driving it with a sufficiently low frequency such that
its parallel wavelength $\lambda_\parallel^-$ is longer than twice the
physical distance $L$ over which the two counterpropagating \Alfven
waves interact, $\lambda_\parallel^-> 2L$. In this case, the length of
the Loop antenna \Alfven wavepacket with which the ASW wave interacts
contains a significant $k_{\parallel }=0$ component, leading to a
nonzero resonant three-wave interaction that transfers energy from the
ASW \Alfven wave to a daughter \Alfven wave with the same parallel
wavenumber (and, thus, the same frequency) but with higher perpendicular
wavenumber.  In Paper III, this concept is demonstrated
quantitatively, and the properties of the nonlinearly generated
daughter \Alfven wave are enumerated.
 
In this paper, we present a detailed analysis of the experimental
measurements to identify unequivocally the nonlinear daughter \Alfven
wave through the verification of the following predicted properties:
\begin{enumerate} 
\item The spatial location of the nonlinear daughter \Alfven  wave should  correspond to the position that can be predicted by the nonlinear term in Eq. (\ref{eq:elsasserpm}). 
\item The daughter \Alfven wave will have the same frequency as the ASW antenna wave signal, $f_D=f_{ASW}$. 
\item The perpendicular wavevector of the  daughter \Alfven wave is given by the vector sum of the perpendicular wavevectors of the ASW and Loop \Alfven waves, $\V{k}_{\perp D} = \V{k}_{\perp ASW} + \V{k}_{\perp Loop}$. 
\item The amplitude of the daughter \Alfven wave agrees with the prediction for a resonant three-wave interaction. 
\end{enumerate} 
 
\section{Experiment} 
 
\subsection{Experimental Setup} 
 
The experiment was conducted in the Large Plasma Device (LaPD) at the Basic Plasma Physics Research Facility at UCLA.\cite{Gekelman1991}  The LaPD was designed specifically to study the \Alfven waves which are relevant to space and astrophysical plasma environments.  Using an indirectly heated barium-oxide coated cathode, the LaPD produces a 16.5 m long, 40-70 cm diameter plasma column with a repitition rate of 1 Hz and a typical discharge lifetime of 10-15 ms.  The experiment took place in an approximately 50\% ionized \cite{Maggs2007} hydrogen plasma.  From a swept Langmuir probe, in conjunction with a microwave interferometer, the density in the measurement region was determined to be $n_e$ = $10^{12} $ cm$^{-3}$ and the electron temperature was $T_e =$ 5 eV.  The background magnetic field was set to $B_0 = $ 800 G, which yields an \Alfven speed of $v_A = 1.75 \times 10^8$ cm$/$s.  From these parameters, the ion cyclotron frequency, $f_{ci} = e B_0/(2 \pi m_i)$, was determined to be 1.2 MHz and the ion sound Larmor radius, $\rho_s = \sqrt{T_e/m_i}/\Omega_i$, was 0.29 cm, where $\Omega_i$ is the angular ion cyclotron frequency.  The ion temperature in the LaPD is typically on the order of $T_i = $ 1 eV, \cite{Palmer2005} although it was not directly measured in this experiment.  From these parameters, the Coulomb logarithm is ln$ \Lambda \simeq 12$ and the electron-ion collison frequency is $\nu_{ei} = 72^{-1/2}\pi^{-3/2}n_ie^4m_e^{-1/2}T_e^{-3/2}\epsilon_0^{-2}$ln$\Lambda \simeq 3$ MHz, so the conditions are moderately collisonal for the \Alfven waves of frequency $f$ generated in this experiment, $f < \nu_{ei}$. 
 
\begin{figure} [!ht] 
\centerline{\includegraphics*[scale=0.475]{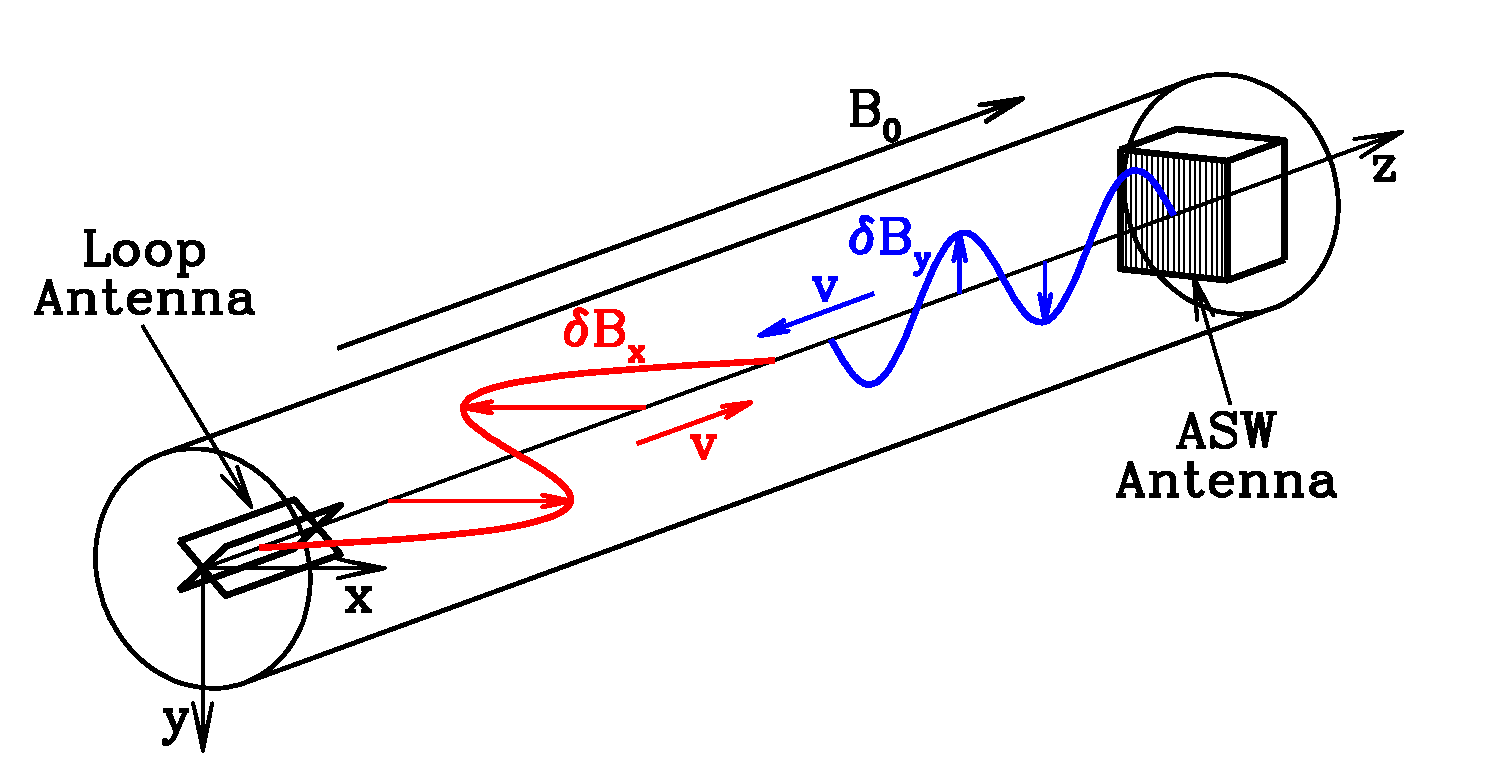}} 
\caption{(Color online) Schematic of the experimental setup for the \Alfven wave turbulence experiment in the LaPD. The ASW antenna generates a small amplutude \Alfven wave (blue line) with a magnetic field polarized in the $\hat{\mathbf{y}}$ direction traveling down the mean magnetic field, $B_0$, and the Loop antenna generates a larger amplitude \Alfven wave (red line) with a magnetic field polarized in the $\hat{\mathbf{x}}$ direction traveling up the mean magnetic field. \cite{Howes2012b} \label{fig:setup}} 
\end{figure} 
 
\begin{figure} [here] 
\centering 
\subfigure [] { 
\label{fig:Iowa_antenna_pic} 
\includegraphics[scale=0.75]{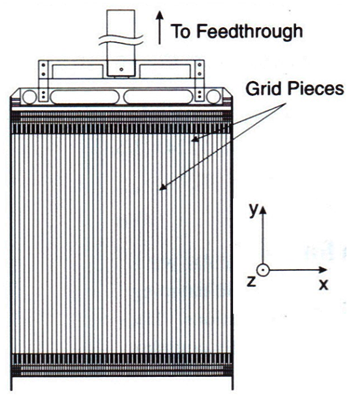}}  
\hspace{7mm} 
\subfigure []{ 
\label{fig:By_Iowa_plot} 
\includegraphics[scale=0.55]{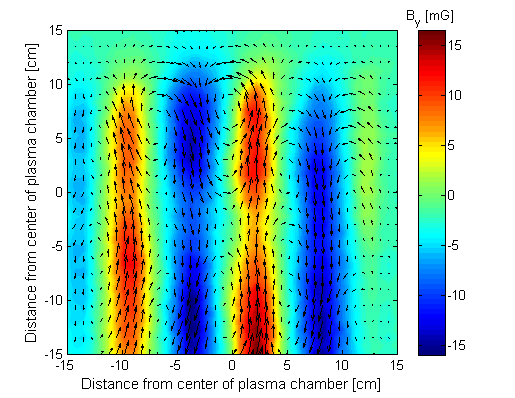}} 
\caption{(Color online) Iowa ASW Antenna (a) diagram and (b) a colormap of the $\delta B_y$ component with vectors indicating the perpendicular component of the magnetic field measured in mG at  $z = 14$ m and $t$ = 8.3 ms after the beginning of the discharge.} 
\label{fig:Iowa_antenna_plots} 
\end{figure}

The counterpropagating wave experiment requires two \Alfven wave antennas placed at either end of the plasma chamber, as shown in Fig. \ref{fig:setup}.\cite{Howes2012b}  The Iowa Arbitrary Spatial Waveform (ASW) antenna\cite{Kletzing2010, Thuecks2009} was placed at $z$ = 15 m, where $ z = 0$ is at the cathode.  This antenna, shown in Fig. \ref{fig:Iowa_antenna_pic}, consists of a set of 48 vertical copper mesh grids of dimension 2.5 cm $\times$ 30.5 cm.  Each element is driven by a separate amplifier which allows the current to be adjusted to a master signal with a multiplicative factor between -1 and 1.  The plane of the mesh grid is oriented perpendicular to the axial magnetic field of the LaPD.  By varying the amplitude of each grid element, we are able to create an arbitrary spatial waveform across the array in the $\hat{x}$ direction with effectively no variation in the $\hat{y}$ direction. Since \Alfven waves have $\delta B_{\parallel} = 0$ and $\nabla \cdot \mathbf{B}= 0$, then $\mathbf{k}_{\perp} \perp \delta \mathbf{B}_{\perp}$.  Therefore, an \Alfven wave with a perpendicular wavevector $\mathbf{k}_{\perp} = k_\perp \hat{\mathbf{x}}$  has a perpendicular magnetic field fluctuation $\delta \mathbf{B}_\perp = \delta B_\perp \hat{\mathbf{y}}$ and, similarly, an \Alfven wave with a perpendicular wavevector $\mathbf{k}_{\perp} = k_\perp \hat{\mathbf{y}}$  has a perpendicular magnetic field fluctuation $\delta \mathbf{B}_\perp = \delta B_\perp \hat{\mathbf{x}}$ .   For this experiment the ASW antenna generated an \Alfven wave (blue line in Fig. \ref{fig:setup}) with a sinusoidal waveform of frequency $f_{ASW} =$ 270 kHz or $\omega / \Omega_{i} \sim 0.22$, a parallel wavelength of $\lambda_{||_{ASW}} = 6.5$ m, and a perpendicular wavevector of $k_x \rho_s \simeq \pm 0.18$, which propagates anti-parallel to the background magnetic field, $\mathbf{B}_0 = B_0 \hat{\mathbf{z}}$.  Note that the ASW antenna will naturally produce both components of the perpendicular wave vector, $\pm k_x$. 
 
\begin{figure} [here] 
\centering 
\subfigure [] { 
\includegraphics[scale=0.65]{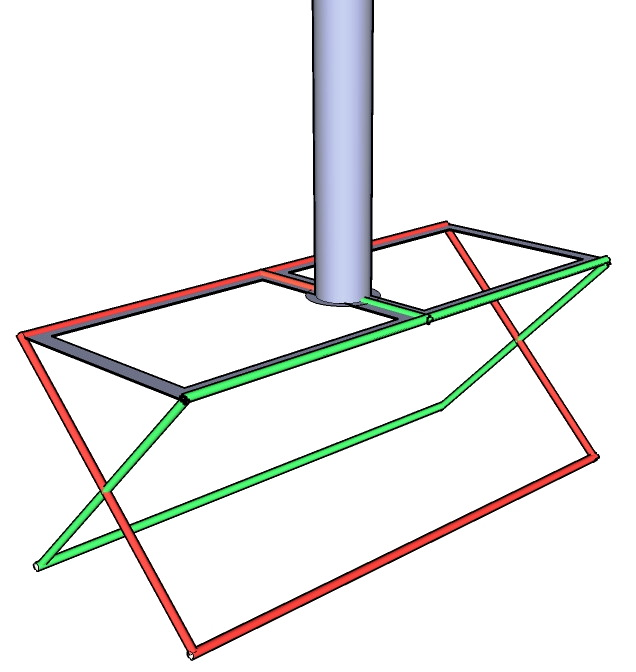} 
\label{fig:UCLA_antenna_pic}} 
\hspace{7mm} 
\subfigure []{ 
\includegraphics[scale=0.55]{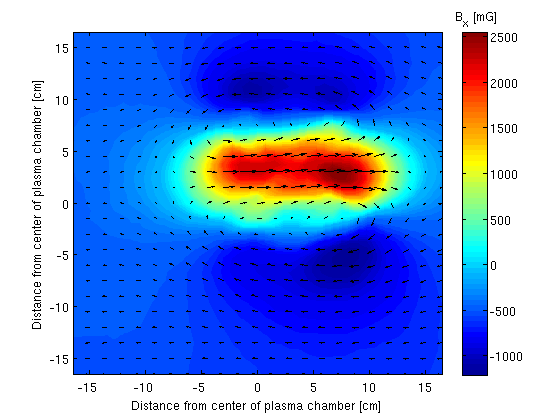} 
\label{fig:Bx_UCLA_plot}} 
\caption{(Color online) UCLA Loop Antenna (a) diagram and (b) a colormap of the $\delta B_x$ component with vectors indicating the perpendicular component of the magnetic field measured in mG at $z = 6.4$ m and $t$ = 8.2 ms after the beginning of the discharge.  The offset in the data for the UCLA Loop antenna was due to a slight biasing issue with the \Elsasser probes.} 
\label{fig:UCLA_antenna_plots} 
\end{figure}

The second antenna used in this experiment was the UCLA Loop antenna,\cite{Auerbach2011} which was placed at $z$ = 4.2 m.  This antenna, shown in Fig. \ref{fig:UCLA_antenna_pic}, consists of two overlapping rectangular loops of dimensions 21.5 cm $\times$ 29.5 cm, which are electrically isolated from each other. By orienting the loops perpendicular to each other and varying the relative phase of the driving signal on each loop, a large amplitude \Alfven wave can be produced with a magnetic field predominately in the $\hat{x}$ direction and a dominant perpendicular wavevecter in the $\hat{y}$ direction with $k_y\rho_s \simeq \pm 0.05$.    For this experiment, the Loop antenna generated an \Alfven wave (red line in Fig. \ref{fig:setup}) with a sinusoidal waveform of  frequency $f_{Loop} =$ 60 kHz, $\omega / \Omega_{i} \sim 0.05$, and a parallel wavelength of $\lambda_{||_{Loop}} = 29.2$ m, which propagates parallel to $\mathbf{B}_0$ .   
 
The perpedicular components of the magnetic field were measured using two \Elsasser probes \cite{Drake2011} placed at $z$ = 6.4 m and $z$ = 14 m. These probes employ two B-dot coils constructed with forty 1.6 mm diameter loops of magnetic wire and oriented such that one is in the $\hat{x}$ plane and one is in the $\hat{y}$ plane.  Measurements were performed over a 30 cm square region, centered on the machine axis, on a grid of locations separated by $\Delta$ = 0.75 cm.  Using an ensemble of plasma discharges, time series with a sample frequency of 25 MHz (temporal resolution $\Delta t = 0.04$ $\mu$s), data were collected at each of the spatial locations in turn with a specified starting time during the shot.  Since the shot to shot variation in the LaPD is modest, averaging over 10 shots per spatial position for the ASW antenna \Alfven wave was sufficient to achieve a RMS noise level of $\sim$ 0.25  mG.  The signal to noise is subsequently greatly enhanced by a spatial Fourier transform because we launch and detect waves that are nearly planar. 
 
\subsection {Experimental Procedure} 
 
The procedure for measuring the magnetic field fluctuations of the two counterpropagating waves follows.  At $t$ = 8.0 ms, where $t$ = 0 s is at the start of the plasma discharge, the Loop antenna launches a wavepacket consisting of 30 wave periods which lasts for around 0.50 ms.  At $t$ = 8.25 ms, the ASW antenna launches a wavepacket consisting of 20 wave periods (duration of 0.074 ms).  Since the Loop antenna launches waves in both directions, that is towards the ASW antenna and towards the cathode, this time delay allows for ample time for the combined direct and reflected waves to reach a steady state before the ASW antenna launches its wave. Perpendicular magnetic field fluctuations are recorded before, during, and after the Loop antenna launches the large amplitude \Alfven wave.  This procedure allows us to measure not only the entire interval when both antennas are turned on, but also the background noise in the plasma.  This experiment was repeated five times with identical timings.  Table \ref{tab:five_trials} shows a summary of the five trials with the levels (max, half, and zero) indicating the amplitude of the magnetic field of each antenna.    
\begin{center} 
\begin{table} 
\caption {Summary of the five experimental trials.}\label{tab:five_trials} 
\begin{tabular} {|c|c|c|} 
\hline 
Trial & ASW Amplitude & Loop Amplitude\\ 
\hline 
1 & maximum & maximum \\ 
\hline 
2 & maximum & zero\\ 
\hline 
3 & zero & maximum \\ 
\hline 
4 & maximum & half \\ 
\hline 
5 & half & maximum \\ 
\hline 
\end{tabular} 
\end{table} 
\end{center} 
 
The measured perpendicular magnetic field fluctuations produced by the ASW \Alfven waves in trial 2 and the Loop antenna waves in trial 3 are shown in Figs. \ref{fig:By_Iowa_plot} and \ref{fig:Bx_UCLA_plot}, respectively.  The colormaps in the figures include a linear interpolation to fill the locations in the plot between the actual measurements.  Since the \Elsasser probe employs a B-dot coil to measure the magnetic field, the measured signals are integrated in time in order to determine $\delta \mathbf{B_{\perp}}$.  The colormaps show $\delta B_y(x, y, t)$ for the ASW antenna at $t = 8.3$ ms and $\delta B_x(x, y, t)$ for the Loop antenna at $t = 8.2$ ms.  The vectors in each figure indicate the total vector $\delta\mathbf{B_{\perp}}$.  The wave generated by ASW antenna has a typical amplitude of 30 mG in $\delta B_y(x, y, t)$ with almost no $\delta B_x(x, y, t)$ contribution, which indicates that the antenna produces a signal with a nearly pure $k_x$ perpendicular wavevector.  On the other hand, the wave produced by the Loop antenna has a dominant $\delta B_x(x, y, t)$ component with a peak-to peak value of around 3500 mG and a small, but not insignifcant, $\delta B_y(x, y, t)$ component,  $\sim$ 400 mG.  The offset in the data for the UCLA Loop antenna is due to a slight biasing issue with the \Elsasser probes.  
 
\section{Analysis of Experimental Measurements} 
 
One simple way of picturing the nonlinear interaction between the counterpropagating \Alfven waves in this experiment is as follows. The large-amplitude Loop antenna \Alfven wave generates a magnetic shear in the axial magnetic field which oscillates at the Loop antenna wave frequency, $f_{Loop}= 60$~kHz. The nonlinear interaction is equivalent to the distortion of the ASW \Alfven wave as it propagates along this sheared magnetic field. The distorted ASW \Alfven wave is simply a linear combination of the initial ASW \Alfven wave and a nonlinearly generated daughter \Alfven wave. It is the primary goal of this experiment to measure and identify definitively this daughter \Alfven wave. 
 
The daughter \Alfven wave measured at the \Elsasser probe is generated by the nonlinear interaction that occurs only over the interaction region between the ASW antenna and the \Elsasser probe, a length of $\Delta z = 8.6$~m, see Fig. 2 in  Paper III.\cite{Howes2013c}  Since $v_A = 1.75 \times 10^8$ cm/s, the time in which the two counterpropagating \Alfven waves may interact nonlinearly is $\Delta t = \Delta z / v_A = 4.9\ \mu$s, less than $1/3$ the Loop antenna wave period, $T_{Loop}=16.7\ \mu$s. Over the time of the nonlinear interaction, the ASW \Alfven wave interacts with only a fraction of the wavelength of the Loop \Alfven wave. The resulting counterpropagating \Alfven wave signal experienced by the ASW \Alfven wave therefore has an effective $k_\parallel =0$ component as shown in Fig. 3 of Paper III. The nonlinear daughter wave generated by the $k_\parallel =0$ component of the Loop \Alfven wave and the ASW \Alfven wave is predicted theoretically to have the following properties: 
 
\begin{enumerate} 
\item The spatial location of the nonlinear daughter wave should correspond to the position that can be predicted by the nonlinear term in Els$\ddot{\mbox{a}}$sser form of the incompressible MHD equations. 
\item The frequency band of the daughter wave will be the centered on the frequency of the ASW antenna wave signal, i.e. $f_D = f_{ASW}$, or $k_{\parallel_D} = k_{\parallel_{ASW}}$. 
\item The nonlinear three-wave interaction should satisfy $\mathbf{k}_{\perp_D} = \mathbf{k}_{\perp_{ASW}}+\mathbf{k}_{\perp_{Loop}}$. 
\item The amplitude of the nonlinear daughter wave should agree with theoretical predictions. 
\end{enumerate}

\begin{figure} [H] 
\centerline{\includegraphics[scale=0.6]{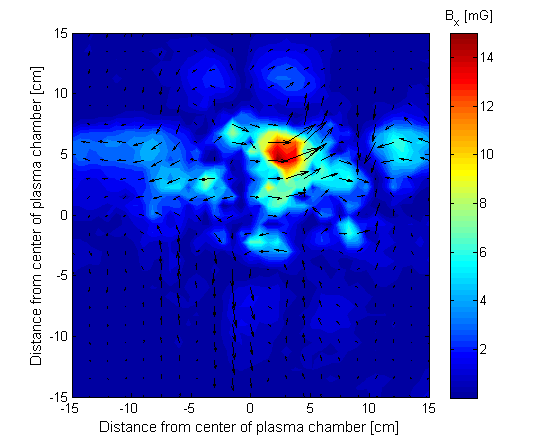}} 
\caption{(Color online) Colormap of the $\delta B_x$ component of the daughter wave with vectors indicating the perpendicular component of the magnetic field at $t$ = 8.30 ms. \label{fig:Bx_diff_plot}} 
\end{figure}

As discussed previously, the \Elsasser probe employs a B-dot coil to measure the $\delta \dot{B}_x(x,y,t)$ and $\delta \dot{B}_y(x,y,t)$ of a fluctuating magnetic field.  From the theory we predict that the daughter wave will have the same frequency as the \Alfven wave produced by the ASW antenna, $f_{ASW} = 270$ kHz.  Since the ASW antenna produces a wave almost exclusively in the $\hat{y}$ direction and  $f_{ASW}$ falls between the third and fourth harmonic of the Loop antenna signal, the only signal measured on $\delta \dot{B}_x(x,y,t)$ component at $f_{ASW}$ should be the daughter wave signal.  Therefore, we first subtract the $\delta \dot{B}_{x_{Loop}}$, data from trial 3, from the data in which both antennas are on and measured at $z = 6.4$ m, trials 1, 4, or 5.  This effectively eliminates the linear contribution from the large amplitude \Alfven wave produced by the Loop antenna.  Next, we integrate this result to obtain $\delta B_{x_D}(x,y,t)$.  From these results we select a time interval in which both waves are measured, 8.25 ms $\le t \le$ 8.32 ms, and Fourier transform this interval in time to obtain $\delta B_{x_D}(x,y,f)$.  Since we expect the daughter wave to have a frequency corresponding to that of the ASW antenna, $f$ = 270 kHz, a bandpass filter is applied to this data to eliminate the frequencies below $f$ = 170 kHz and above $f$ = 370 kHz.  Finally, the resulting data sequence is inverse Fourier transfromed back into the time domain, $\delta B_{x_D}(x,y,t)$.  The results of this analysis are shown at $t$ = 8.30 ms in Fig. \ref{fig:Bx_diff_plot}.  
  
\subsection{Spatial Localization} 
 
We can use the nonlinear term in Eq.~(\ref{eq:elsasserpm}) to predict the position where the nonlinear daughter wave will appear in the experiment. This nonlinear term describing the distortion of the ASW \Alfven wave by the Loop \Alfven wave is given by $\V{z}^-_{Loop} \cdot \nabla \V{z}^+_{ASW}$. The  eigenfunction  for an \Alfven wave traveling up the magnetic field determines that $\V{v}_\perp = -\delta \V{B}_\perp/\sqrt{\mu_0 \rho_0}$, so the \Elsasser field for the Loop  \Alfven wave can be expressed simply in terms of its magnetic perturbation 
$\V{z}^-_{Loop} = - 2\delta \V{B}_{\perp Loop}/ \sqrt{\mu_0 \rho_0}$. Similarly, the eigenfunction  for an \Alfven wave traveling down the magnetic field is given by $\V{v}_\perp = \delta 
\V{B}_\perp/\sqrt{\mu_0 \rho_0}$, so that  $\V{z}^+_{ASW} = 2\delta \V{B}_{\perp ASW}/ \sqrt{\mu_0 \rho_0}$. Note that this eigenfunction is correct not only in the MHD limit of 
strong collisionality and large scales, $k \rho_s \ll 1$, but also in the limit appropriate for the LaPD experiment of moderate collisionality and large scales.\cite{Schekochihin2009, Howes2013a, Howes2013b, Howes2013c} 
 
\begin{figure} [top] 
\centerline{\includegraphics[scale=0.6]{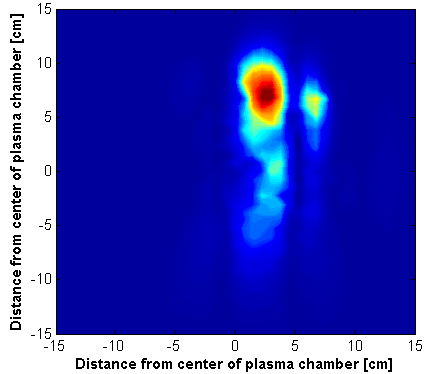}} 
\caption{(Color online) Colormap of the predicted coupling between the Loop and ASW antennas.  The dark spot at ($x$ = 3 cm, $y$ = 7 cm) indicates the position in the plane where the nonlinear effect is predicted to have the largest amplitude. \label{fig:Dot_product_plot}} 
\end{figure}

Since the variation of the magnetic field of the ASW \Alfven wave is only in the $x$-direction, this nonlinear term simplifies to  
\begin{equation} 
\delta B_{x Loop} \frac{\partial\delta B_{y ASW}}{\partial x}, 
\end{equation} 
where we have dropped constant factors. The daughter \Alfven wave is generated by this term,\cite{Note1} which predicts that the magnetic field for the daughter \Alfven wave will be maximum at the spatial position where the Loop antenna's magnetic field, $\delta B_{x Loop}$, is largest and the gradient of the ASW antenna's magnetic field, $ \partial/\partial x( \delta B_{y ASW})$, is largest. We can employ the magnetic field patterns from the single-antenna runs, shown in  Figs.~\ref{fig:By_Iowa_plot} and \ref{fig:Bx_UCLA_plot}, to compute this term to predict the position of the maximum amplitude of the nonlinearly generated daughter \Alfven wave, as shown in Fig.~\ref{fig:Dot_product_plot}.  The result of this very simple prediction agrees well with the measurement of the daughter wave, shown in Fig.~\ref{fig:Bx_diff_plot}, which has a maximum value of 14 mG at position $\delta B_x$ ($x$ = 3 cm, $y$ = 5 cm). This agreement corresponds to the first of our listed predictions for the properties of the daughter \Alfven wave generated by the nonlinear interaction between the two counterpropagating \Alfven waves launched in the experiment. 
 
\subsection{Frequency selection} 
 
\begin{figure} [H] 
\centerline{\includegraphics [scale=0.7]{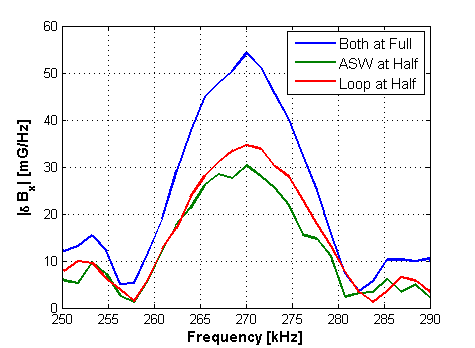}} 
\caption{(Color online) Plot of the amplitude of the daughter wave in frequency where the maximum of the nonlinear signal occurs, $\delta B_x$ ($x$ = 3 cm, $y$ = 5 cm, $f$).  We show the results for when both antennas are at full power (blue), the ASW antenna is at half power and the Loop antenna is at full power (green), and when the Loop antenna is at half and the ASW antenna is at full (red).\label{fig:Halved_amplitude_plot}} 
\end{figure}

The second property predicted for the daughter wave is that the frequency of the daughter wave is the same as the frequency of the ASW antenna wave, $f_D = f_{ASW}$ or equivalently $k_{\parallel_D} = k_{\parallel_{ASW}}$.  Since the ASW wave antenna has a magnetic field predominantly in the $\hat{y}$ direction, one way to distinguish the daughter wave from the ASW wave is to look at the $\delta B_x$ component of the waves in the frequency domain.  Because the daughter wave is a nonlinear effect, the amplitude should vary as the product of the primary wave amplitudes, as shown by Eq. 36 in Paper I.\cite{Howes2013a}  Thus, if the amplitude of either antenna signal is reduced by half, the amplitude of the daughter wave should also be reduced by half.  If the signal at 270 kHz was related to one of the primary waves, this would not occur.  Thus by observing what occurs when the Loop and ASW antenna amplitudes are individually decreased by half, we can look for the presence of a nonlineraly generated daughter wave.  In Fig. \ref{fig:Halved_amplitude_plot}, we show the daughter wave signal (blue) at 270 kHz when both antennas are at full power and at position $\delta B_x$ ($x$ = 3 cm, $y$ = 5 cm, $f$), which is where the maximum of the nonlinear effect occurred as shown in Fig. \ref{fig:Bx_diff_plot}.  Fig. \ref{fig:Halved_amplitude_plot} also shows, at the same axial position, results for trial 4, when the Loop antenna signal is turned down to half power and the ASW antenna is kept at full power (red), and trial 5, when the ASW antenna is turned to half power and the Loop antenna is at full (green).  These results clearly demonstrate that the daughter wave signal decreases by the same amount ($\sim$ 40\%) when either of the antenna amplitudes is decreased.  Note that when the two antennas are at full power, the antenna coupling into the plasma is starting to saturate, and thus the response is not perfectly linear. 
 
\subsection{Wave number selection} 
 
\begin{figure*} [!ht] 
\centering 
\subfigure [] { 
\includegraphics[scale=0.5]{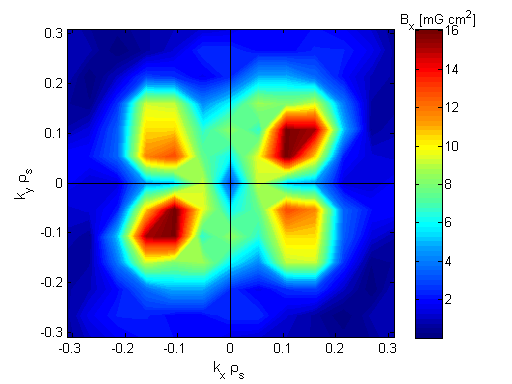}}  
\subfigure [] { 
\includegraphics[scale=0.5]{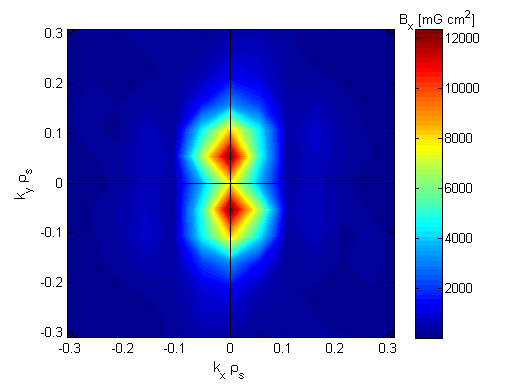}}  
\subfigure []{ 
\includegraphics[scale=0.5]{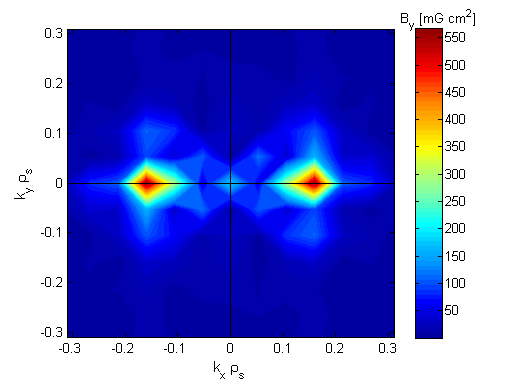}} 
\hspace{0.2cm} 
\subfigure []{ 
\includegraphics[scale=0.5]{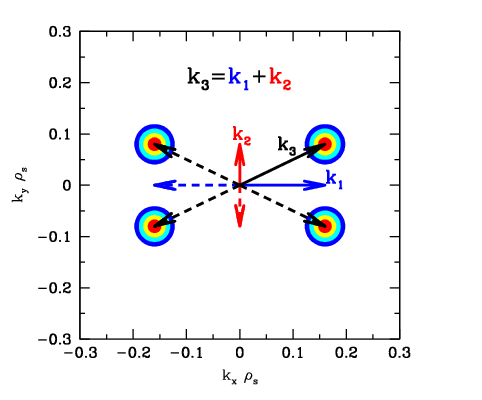}} 
\caption{(Color online) Contour plots of the two-dimensional Fourier power spectrum of the (a) $\delta B_x$ component of the daughter signal, (b) the $\delta B_x$ component of the Loop signal, (c) the $\delta B_y$ component of the ASW antenna signal, and (d) is the diagram of the perpendicular wavevectors for the Loop antenna (red line) and the ASW antenna (blue line).  The daughter wave should be a vector sum of the type, $\mathbf{k}_3 = \mathbf{k}_1 + \mathbf{k}_2$.  The bullseyes indicate the predicted values for the daughter wave. \cite{Howes2012b}} 
\label{fig:kx_ky_plots} 
\end{figure*}

The third prediction for the daughter wave signal is that $\mathbf{k}_{\perp_D} = \mathbf{k}_{\perp_{ASW}}+\mathbf{k}_{\perp_{Loop}}$.  To obtain the spatial Fourier transform in the perpendicular plane for each antenna signal and the daughter wave signal, we Fourier transform the data shown in Figs. \ref{fig:By_Iowa_plot}, \ref{fig:Bx_UCLA_plot}, and \ref{fig:Bx_diff_plot} in both the $\hat{x}$ and $\hat{y}$ directions, which yields $\delta \mathbf{B}_{x}(k_x,k_y,t)$ for the Loop antenna signal and  $\delta \mathbf{B}_{y}(k_x,k_y,t)$ for the ASW antenna.  We present these results in Fig. \ref{fig:kx_ky_plots} at $t$ = 8.30 ms.\cite{Howes2012b}  In Fig. \ref{fig:kx_ky_plots} (b) we show the two-dimensional Fourier transform of the Loop antenna data in Fig. \ref{fig:Bx_UCLA_plot}, which shows that the the $\delta B_x$ component of the Loop antenna has a perpendicular wavevector of $k_y \rho_s = \pm 0.06$.  Fig. \ref{fig:kx_ky_plots} (c) shows the Fourier transform of the $\delta B_y (k_x, k_y, t)$ of the ASW antenna signal in Fig. \ref{fig:By_Iowa_plot}, clearly indicating that the perpendicular wavevector for the $\delta B_y$ component of the ASW antenna is $k_x \rho_s = \pm 0.16$.  This value is close to the expected value of $k_x \rho_s \simeq \pm 0.2$ (see Sec. III A).  The small discrpency in these two values is because the density profile is not perfectly flat, which can produce a small shift in the $k_\perp$ of an \Alfven wave.  The resulting daughter wave, $\delta B_x(k_x, k_y, t)$, is shown in Fig. \ref{fig:kx_ky_plots} (a), which was taken from the Fourier transform of Fig. \ref{fig:Bx_diff_plot}.  We can clearly see that the daughter wave signal is a sum of the ASW and Loop antenna wavenumbers,  $\mathbf{k}_{\perp_D} = + (\mathbf{k}_{\perp_{ASW}} \pm \mathbf{k}_{\perp_{Loop}})$ and  $\mathbf{k}_{\perp_D} = - (\mathbf{k}_{\perp_{ASW}} \pm \mathbf{k}_{\perp_{Loop}})$.  A diagram of the three wave interaction process is shown in panel (d).  Here $\mathbf{k}_1$, in blue, indicates the perpendicular wavevector contributions of the $\delta B_y$ of the ASW antenna.  $\mathbf{k}_2$, shown in red, is the perpendicular wavevector produced by the $\delta B_x$ of the Loop antenna.  Note that both antennas produce a pair of wavevectors with $\pm \mathbf{k}_{1,2}$. The daughter wave should be a vector sum of the type, $\mathbf{k}_3 = \mathbf{k}_1 + \mathbf{k}_2$.  The bullseyes indicate the predicted values for the daughter wave.  These predictions align  well with experimental results in panel (a).   
 
\subsection{Amplitude of Daughter wave} 
 
A final line of evidence that the signal measured in the experiment is the nonlinearly generated daughter \Alfven wave is to compare the measured magnitude of the signal to the theoretical prediction. An asymptotic solution for the nonlinear evolution of the interaction between counterpropagating \Alfven waves has been derived in Paper I \cite{Howes2013a} of this series.   The second-order nonlinear solution for two counterpropagating \Alfven waves with equal $k_\perp$ and $k_\parallel$ is given by Eq. 36 in Paper I.  Therefore, we estimate that the nonlinear daughter wave in our case will have an amplitude  
 
\begin{equation} 
\frac{|\V{B}_{\perp 2}|}{B_0} =  \frac{  z_+ z_-}{ 16 v_A^2}  
\frac{  k_\perp }{k_\parallel} 
\end{equation} 
 
Using the eigenfunction for an \Alfven wave, ${u_\perp}/{v_A} = \pm{B_\perp}/{B_0}$, the magnitude of the \Elsasser variables $z^\pm$ are related to the magnetic field perturbation by $ z_\pm / v_A= 2{B}^\pm_{\perp}/B_0$. 
 
We can therefore express the nonlinear daughter wave magnetic amplitude $\delta {B}_{\perp D}$ in terms of the loop wave amplitude $\delta {B}_{\perp L}$ and the ASW wave amplitude $\delta {B}_{\perp A}$, 
 
\begin{equation} 
\frac{\delta B_{\perp D}}{B_0} =  \frac{1}{4 }\frac{{\delta B}_{\perp L} }{B_0} \frac{{\delta B}_{\perp A} }{B_0}  \frac{  k_\perp }{k_\parallel} 
\end{equation} 
 
The normalized amplitudes in this expression correspond to the amplitudes of the Fourier coefficients in Fig. \ref{fig:kx_ky_plots}, corresponding to $B_{\perp D}=16$~mG~cm$^2$, $B_{\perp L}=12000$~mG~cm$^2$, and $B_{\perp A}=550$~mG~cm$^2$. These values must be normalized to an appropriate value of $B_0$, so we use the ratio of the maximum Loop antenna magnetic field magnitude from Fig.~\ref{fig:Bx_UCLA_plot} divided by the axial magnetic field to estimate the normalized value for $B_{\perp L}/B_0 = 1.5\mbox{G}/800\mbox{ G}= 1.9 \times 10^{-3}$.  Using this value, we can estimate the other normalized values as $B_{\perp A}/B_0=8.7 \times 10^{-5}$ and $B_{\perp D}/B_0=2.5 \times 10^{-6}$.  To compute the ratio $ k_\perp /k_\parallel$ for the ASW \Alfven wave, we take $\lambda_\parallel=v_A /f= 1.75 \times 10^{8} \mbox{ cm s}^{-1}/2.7\times 10^5 \mbox{ Hz}=648$~cm, and $\lambda_\perp=10.16$~cm.  This leads $ k_\perp /k_\parallel= \lambda_\parallel/\lambda_\perp=64$. 
 
Substituting these values into Eq. 36 of Paper I, we obtain a predicted normalized amplitude of $B_{\perp D}/B_0=2.6 \times 10^{-6}$, in striking agreement with the measured normalized value of $B_{\perp D}/B_0=2.5 \times 10^{-6}$.  This calculation indicates that the amplitude of the nonlinearly generated wave agrees extremely well with the predictions from analytical theory.

\section{Conclusions} 
 
In this paper we present the first experimental verification of the properties of the nonlinear interaction between two counterpropagating \Alfven waves as derived from incompressible MHD.  We have confirmed that the nonlinear interaction between the Loop antenna and the ASW antenna is well described by the nonlinear term in the \Elsasser equation, Eq.~(\ref{eq:elsasserpm}).  In this case, the maximum for the nonlinear signal will appear at the location where both the magnitude of the Loop antenna and the gradient of the ASW antenna are largest.  This was clearly evident in the comparison of the measured spatial plot of the daughter wave, Fig. \ref{fig:Bx_diff_plot}, and the colormap of the predicted location, Fig. \ref{fig:Dot_product_plot}. From Fig. \ref{fig:Halved_amplitude_plot}, we saw that the daughter wave has a peak at the expected frequency of 270 kHz, which corresponds to the frequency of the ASW antenna wave signal.  In Fig. \ref{fig:kx_ky_plots}, we observed that the perpendicular structure of the daughter wave is dominated by four wavevectors corresponding to  $\mathbf{k}_{\perp_D} = \mathbf{k}_{\perp_{ASW}}+\mathbf{k}_{\perp_{Loop}}$.  Since the measured daughter wave signal satisfies the theoretical predictions, we conclude that we have measured the nonlinear interaction between two counterpropagating \Alfven waves.  This evidence supports the use of such idealized models, as discussed in Paper I,\cite{Howes2013a} for weakly collisional plasmas which are relevant to various space and astrophysical plasma environments.   It is important to note that this procedure, of subtracting the parent wave signals and bandpass filtering the data, is not required to see the nonlinear daughter wave signal since the three waves are in different locations in the $k$-plane and vector orientation of $\delta B_\perp$ in the plane is different.  Although the parent \Alfven wave signals are much greater in magnitude then the daughter wave signals, the contributions of the parent waves to the daughter signal at the particular $k$-plane location, vector orientation, and frequency is very modest and well resolved given the RMS noise level of $\sim 0.25$ mG.

 
%
%
 
%
 
\begin{acknowledgments} 
Funding for this project was provided by NSF PHY-10033446, NSF CAREER AGS-1054061, NSF CAREER PHY-0547572, and NASA NNX10AC91G.  The experiments presented here were conducted at the Basic Plasma Science Facility at UCLA, which is funded in part by the U.S. Department of Energy and the NSF. 
\end{acknowledgments} 
 
%
 
\end{document}